\documentclass[twocolumn]{aastex631}

\usepackage{newtxtext,newtxmath,color}
\usepackage{amsmath, graphics, bm, graphicx}
\allowdisplaybreaks

\begin{document}

\title[Cavitation and Chondrites]
{CAVITATING BUBBLES IN CONDENSING GAS\\AS A MEANS OF FORMING CLUMPS, CHONDRITES, AND PLANETESIMALS}

\author[0000-0002-6246-2310]{E.~Chiang}
\affiliation{
Department of Astronomy, University of California, Berkeley, CA 94720-3411, USA \\
}
\affiliation{
Department of Earth and Planetary Science, University of California, Berkeley, CA 94720-4767, USA
}

\date{Accepted Sep 3 2024 to ApJ Letters
}



\begin{abstract}
Vaporized metal, silicates, and ices on the verge of re-condensing into solid or liquid particles appear in many contexts: behind shocks, in impact ejecta, and within the atmospheres and outflows of stars, disks, planets, and minor bodies. We speculate that a condensing gas might fragment, forming overdensities within relative voids, from a radiation-condensation instability. Seeded with small thermal fluctuations, a condensible gas will exhibit spatial variations in the density of particle condensates. Regions of higher particle density may radiate more, cooling faster. Faster cooling leads to still more condensation, lowering the local pressure. Regions undergoing runaway condensation may collapse under the pressure of their less condensed surroundings. Particle condensates will compactify with collapsing regions, into overdense clumps or macroscopic solids (planetesimals). As a first step toward realizing this hypothetical instability, we calculate the evolution of a small volume of condensing silicate vapor---a spherical test ``bubble'' embedded in a background medium whose pressure and radiation field are assumed fixed for simplicity. Such a bubble  condenses and collapses upon radiating its latent heat to the background, assuming its energy loss is not stopped by background irradiation. Collapse speeds can range up to sonic, similar to cavitation in terrestrial settings. Adding a non-condensible gas like hydrogen to the bubble stalls the collapse. We discuss whether cavitation can provide a way for mm-sized chondrules and refractory solids to assemble into meteorite parent bodies, focusing on CB/CH chondrites whose constituents likely condensed from silicate/metal vapor released from the most energetic asteroid collisions.
\end{abstract}



\section{Introduction}
\label{sec:intro}
Ubiquitous in meteorites, 
the millimeter-sized igneous spheres that are chondrules 
pose two problems: how to melt them in space, and how to assemble them into asteroids.

The first problem, chondrule melting, has had a number of proposed solutions. These include heating from shock waves in the solar nebula, as driven by planetesimals (e.g.~\citealt{morris_etal_2012,boley_etal_2013}; \citealt{stammler_dullemond_2014}), planets (e.g.~\citealt{bodenan_etal_2020}), or disk self-gravity (e.g. \citealt{desch_connolly_2002}; \citealt{boss_durisen_2005}; \citealt{boley_durisen_2008}). Alternatively, circumstellar solids that drift too close to the magnetospheric cavities of their young host stars may be heated by stellar flares (\citealt{shu_etal_1996, shu_etal_1997, shu_etal_2001}). Nebular lightning (electrons accelerated in a breakdown electric field) has also been proposed as an energy source (e.g.~\citealt{muranushi_2010,johansen_okuzumi_2018,hashimoto_nakano_2021,kaneko_etal_2023}).

Still another means of melting is asteroid collisions. There are many flavors of collisional scenarios: chondrules as melt droplets splashed out from planetesimals already molten from $^{26}$Al (\citealt{asphaug_etal_2011}; \citealt{dullemond_etal_2014,dullemond_etal_2016}); chondrules as melt jetted from the impact site between solid planetesimals (\citealt{johnson_melosh_2015}; \citealt{wakita_etal_2017,wakita_etal_2021}); or chondrules as solids thermally processed by impact-generated vapor (\citealt{stewart_lpsc6, stewart_lpsc7}; \citealt{stewart_lpsc3}; \citealt{carter_stewart_2020}) and radiation (\citealt{choksi_etal_2021}). The impact models are not without their problems, among them avoiding elemental differentiation in liquid/molten asteroids as most chondrules are undifferentiated (e.g.~\citealt{lichtenberg_etal_2018}), and the low efficiencies with which colliding solids produce vapor ($\lesssim$20\%; \citealt{watt_etal_2021}) and melt ($\lesssim$7\%  for impacts with dunite, and orders of magnitude less for dunite mixed with ice;
\citealt{cashion_etal_2022}). Nevertheless, a good case can be made that the subset of chondrites classified CB or CH (\citealt{weisberg_etal_2006}; \citealt{jacquet_2022}) have a collisional origin: their petrologic textures (\citealt{campbell_etal_2002}; \citealt{krot_etal_2007}; \citealt{ivanova_etal_2008}), elemental abundances (\citealt{fedkin_etal_2015}), and thermal histories (\citealt{choksi_etal_2021}) can be reproduced by vapor condensation and radiative cooling in high-velocity impacts between asteroids 10--100 km in radius.
CB chondrites contain a special mix of Fe/Ni metal  nodules and grains, and silicate chondrules unusually devoid of metal. This composition is interpreted to arise from the vaporization of an asteroid formerly differentiated into a metal core and a silicate mantle, and its subsequent recondensation into chondrules and metallic particles (\citealt{oulton_etal_2016,johnson_etal_2016}).

The second problem---collecting chondrules, after they have been heated, into chondrite parent bodies---seems to have received less attention. The problem would seem acute for CB/CH chondrules, which are understood to form from vaporizing collisions. The sign seems wrong---debris from hypervelocity impacts is ejected at speeds comparable to the sound speed of rock vapor, on the order of km/s, readily escaping the gravity wells of bodies 100 km in size or smaller. Even if some chondrule debris were to fall back onto the post-collision remains (e.g.~\citealt{morris_etal_2015}), or be accreted onto other asteroids (e.g.~\citealt{johnson_melosh_2015}), accretion velocities tend to be so high as to shatter chondrules on impact,\footnote{Although 
``chondrule-like'' fragments $< 30$ $\mu$m in size have been collected from the comet 81P/Wild-2 (\citealt{nakamura_etal_2008}) and the asteroid Ryugu (\citealt{nakashima_etal_2023}). Such trace micro-chondrules may be the remains of chondrules that shattered upon impacting the surfaces of asteroids and comet progenitors (\citealt{choksi_etal_2021}). In-situ collisions, and the heating and re-accretion of solids resulting therefrom, may (contrary to prevailing wisdom; cf.~\citealt{brownlee_etal_2012}) obviate the need for large-scale transport of solids from heated regions near the proto-Sun.} 
and to defeat gravitational focussing and render accretion improbable (\citealt{choksi_etal_2021}). In reality, chondrules must have agglomerated gently enough to remain intact, and with remarkable efficiency to fill the majority of the volumes of chondritic meteorites. In the case of CB/CH chondrites, the volume-filling fraction of chondrules and metal nodules is $\gtrsim 95\%$ (e.g.~\citealt{jacquet_2022}). To achieve such efficiency, the mechanism of aggregating chondrules was likely part and parcel of the same mechanism that melted them. That is, between the heating and assembly phases, chondrules must not have dallied and mixed unduly
with other nebular solids.\footnote{Though some mixing is necessary, as the chondrule + Fe/Ni metal volume-filling fraction is never 100\% in any chondrite. Typical chondrites are a sedimentary-like amalgamation of chondrules, Fe/Ni metal, calcium-aluminium inclusions (CAIs), and unheated matrix grains, components with different chemical and thermal histories (e.g.~\citealt{connolly_jones_review}; for a review of CB/CH constituents, see \citealt{krot_etal_2002}). Furthermore, the chondrules within a given chondrite may have different ages. CB chondrites contain not only young chondrules, but also CAIs that are 4--5 Myr older \citep{krot_etal_2005, wolfer_etal_2023}.}

Despite appearances, a hot expanding vapor cloud generated from an asteroid collision may yet be conducive to planetesimal (re)formation, as proposed by \citet{stewart_lpsc6,stewart_lpsc7} and \citet{stewart_lpsc3}. Their numerical simulation of an impact vapor plume shows that its expansion can be halted and reversed: cooling and condensation create a pressure void which collapses under the background pressure from nebular H$_2$. The simulated implosion in \citet{stewart_lpsc6} is arguably exaggerated, as it assumes the background H$_2$ is stationary with respect to the target asteroid; the roughly symmetric expansion of ejecta into a static medium allows for its subsequent, roughly symmetric collapse.\footnote{Technically axisymmetric about the velocity vector of the impactor relative to the target. Debris is strewn along an axis extending from the impact point to what is left of the impactor barreling through the assumed static nebula. Vapor expands outward from this axis, in a cone-shaped plume with the impactor at its vertex.} A more realistic set-up would account for H$_2$ gas moving relative to both target and impactor, i.e., the km/s nebular headwind due to the colliders' non-circular orbits, which are necessarily eccentric and/or inclined to yield a vaporizing collision in the first place. The hydrogen headwind would be expected to sweep through and flush away the condensed remains of the impact plume, rather than implode the plume from all directions (\citealt{choksi_etal_2021}).

We consider here the possibility that condensation-induced collapse might still occur within the plume on smaller scales. \citet{stewart_lpsc3} note ``eddies'' and ``clumping of the entrained condensates'' in their simulation. It is not hard to imagine how a gas might be susceptible to heterogenous condensation and localized collapse, even without an external confining nebula. Consider a purely condensible gas like silicate vapor, with no H$_2$ anywhere, on the verge of condensing. Because the saturation vapor pressure is exponentially sensitive to temperature, small variations in temperature across the gas yield larger-amplitude variations in the degree of saturation. Cooler regions saturate first. The resultant condensations may run away: condensed liquids/solids (``dust particles'') are efficient emitters of thermal radiation,\footnote{The simulation of \citet{stewart_lpsc6,stewart_lpsc7} and \citet{stewart_lpsc3} did not account for radiation.} cooling the condensing regions still further and accelerating condensation. These pockets of high condensation and low pressure will be compressed by surrounding lower condensation, higher pressure gas, potentially achieving densities high enough to self-gravitate (e.g.~\citealt{alexander_etal_2008}). Thus, assuming a seed field of thermal fluctuations, the condensing medium may fragment into clumps of condensed particles---on their way, perhaps, to becoming planetesimals. In the context of an asteroid collision, melt droplets within the hot vapor plume could be compactified and reconstituted into chondrites or chondrite progenitors, before being swept downstream by the nebular headwind.

Here we explore condensation-induced collapse with some simple calculations. Within a condensing gas we consider a small test region or ``bubble'', and study how the bubble leaks energy by radiation and compresses under the pressure of the surrounding medium. We assume for simplicity that the medium exterior to the bubble has given and fixed properties (in particular pressure and radiation field). Not following the evolution of the background medium, or writing down its governing equations, prevents us from studying its thermal stability (cf.~\citealt{field_1965}), and testing whether our hypothesized radiation-condensation instability is real.  
Instead, in this preliminary study, we rig up bubbles that can collapse, identify the circumstances under which they do, and describe quantitatively how the collapse unfolds, establishing orders of magnitude.  Our assumption that background properties do not change may not be unrealistic insofar as bubbles can collapse quickly---we will see that evolution times can be measured in seconds to minutes. The physics studied here relates to pressure oscillations of air bubbles in water (\citealt{minnaert_1933}), as well as cavitation: the formation and collapse of vapor bubbles in low-pressure liquids (see, e.g., \citealt{thorne_blandford_2017}). In terrestrial examples of cavitation, collapse speeds can be sonic and generate shock waves; ram pressures from imploding bubbles can damage fast-moving machinery immersed in liquid (watercraft propellers, gear pumps), or disable the prey of fast-striking marine animals (pistol shrimp, mantis shrimp). In the context of chondrule formation, cavitation has also been invoked to limit the sizes of molten droplets experiencing gas drag (\citealt{miura_nakamoto_2007}), and to create chondrules from the popping of vapor bubbles off initially solid bodies heated to boiling (\citealt{nakano_hashimoto_2020}). These works consider cavitation on mm-cm scales; our work asks whether a different kind of cavitation can operate on much larger, super-km scales, not to create chondrules, but to assemble them into parent bodies.

In section \ref{sec:model} we write down and solve the equations governing the dynamics and thermodynamics of condensing spherical bubbles in a fixed background. Though we do not spatially resolve the bubbles, and use expressions that are at best accurate to order-of-magnitude, we can follow the evolution of bubbles across a wide range of scales and circumstances, and also assess what size solid particles can be entrained in collapsing gas. In section \ref{sec:gas} we consider how mixing a non-condensible gas (hydrogen) into the condensing bubble can qualitatively change outcomes. Section \ref{sec:sum} summarizes and connects our results to problems of chondrite formation and structure formation generally.


\vspace{0.75in}
\section{Bubbles of Wholly Condensible Vapor}
\label{sec:model}
To re-cap the big picture motivated in the Introduction: we consider how a portion of a condensible medium evolves under perturbation. The medium can be drawn from any chondrule-forming event, staged concurrently with chondrule formation or in its immediate aftermath---e.g., the protostellar nebula downstream of a shock, or the debris cloud from a hypervelocity impact between asteroids---any region containing hot rock vapor and either freshly melted chondrules or soon-to-be-condensed chondrules, that subsequently cools by expansion and radiation. We analyze a piece of the medium, a ``bubble'', whose properties are for whatever reason slightly different from its surroundings, and ask how the bubble interacts with the background, focusing on how the bubble may lose energy to the background through radiation, and be crushed into a smaller volume by the background pressure. While we evolve the bubble properties in time, we freeze the background---this is done for simplicity, and is justifiable insofar as the bubble might evolve on a timescale faster than the background evolves. This timescale requirement is not necessarily satisfied, and we will butt up against it. In the case of a background medium drawn from the vapor plume from an asteroid impact,  \citet{choksi_etal_2021} showed that thermal histories of impact plumes produced by colliding asteroids 10-100 km in radius (these reproduce empirical chondrule cooling rates of 100-1000 K/hr) unfold over hours, possibly stretching to days, and create plumes on the order of 10000 km across.  We are asking here how such impact plumes might fragment on smaller scales, on timescales shorter than days.

For this first set of calculations (section \ref{sec:model}), we assume the medium is composed purely of silicates and metals, a fraction of which is in the vapor phase and capable of condensing wholly into solid/liquid ``dust'' particles (we do not distinguish between solid and liquid phases and hereafter refer to dust particles as ``solid''). The medium has uniform pressure $P_{\rm bkg}$ and temperature $T_{\rm bkg}$. Out of this background we carve a test sphere (bubble) of radius $R$, temperature $T$, and pressure $P$.

The equations governing the evolution of the bubble are described in section \ref{sec:basic}. Results for a fiducial set of parameters are presented in section \ref{sec:fiducial}, and parameter space explored in section \ref{sec:param}. In section \ref{sec:drag}, we consider what size solids may be aerodynamically dragged with the bubble, and what particle-particle relative velocities may be. Section \ref{sec:gas} introduces a non-condensible gas like nebular hydrogen to the problem.

\subsection{Equations solved}\label{sec:basic}

The gas in the bubble is treated as ideal:
\begin{align}
P = \frac{\rho_{\rm gas} k T}{\mu m_{\rm H}}
\end{align}
where $\rho_{\rm gas}$ is the gas mass density ($< \rho$, the total mass density in gas + solids), $k$ is Boltzmann's constant, $\mu= 30$ is the mean molecular weight, and $m_{\rm H}$ is the atomic mass unit.

While the total bubble mass $M$ is constant, the bubble mass in solid form, $M_{\rm solid}$, can evolve. If the bubble is under-saturated, i.e.~if $P < P_{\rm sat}(T)$,  where $P_{\rm sat}$ is the 
saturation pressure of vapor over rock, then the bubble remains purely vapor:
\begin{equation}
M_{\rm solid} = 0 \,\,\,\,\,\,\,\,\,\,\,\,\,\,\,\,\,\,\,\,\,\,\,\,\,\,\,(P < P_{\rm sat}) \,.
\end{equation}
Otherwise the bubble is saturated and solids  condense out:
\begin{equation}
M_{\rm solid} = M - \rho_{\rm sat}(T) V \,\,\,\,\,\,\,\,(P = P_{\rm sat}) \label{eq:m_solid}
\end{equation}
where $V = 4\pi R^3/3$ is the bubble volume and $\rho_{\rm sat} = P_{\rm sat} \mu m_{\rm H} / (kT)$ is the saturated gas density. We use here the saturation vapor pressure curve of molten bulk silicate Earth (``BSE''), which has a chemical composition similar to that of olivine-rich chondrites (\citealt{fegley_schaefer_2012}; see also \citealt{visscher_fegley_2013}):
\begin{equation}
\log_{10}\left(\frac{P_{\rm sat}}{\rm bars}\right) = -30.6757 - \frac{8228.146\,\,{\rm K}}{T}+\, 9.3974\log_{10}\left(\frac{T}{{\rm K}}\right) \,.
    \label{eq:psat}
\end{equation}
Modeling the medium as a single, continuously condensible species ignores the reality that different minerals condense out at different temperatures; indeed, this discrete condensation sequence (e.g. \citealt{fegley_etal_2023}, their Table 32) is critical to explaining the segregated components of Fe/Ni metal, and metal-free silicate chondrules, in CB/CH chondrites (e.g. \citealt{campbell_etal_2002,fedkin_etal_2015,oulton_etal_2016}). Depending on the composition of rock melt, saturation vapor pressures can vary by 1--2 orders of magnitude at a given temperature. We adopt equation (\ref{eq:psat}) for simplicity and to capture a key feature of condensation-induced collapse, namely the exponential sensitivity of pressure to temperature (Clausius-Clapeyron).

Equation (\ref{eq:m_solid}) follows from our assumption that the bubble can never be super-saturated (e.g.~\citealt{zr,melosh_1989}). The ``excess'' mass---what would super-saturate the bubble if it were a gas\footnote{In practice we determine whether the bubble is under-saturated or saturated by calculating a hypothetical pressure if all of mass $M$ were in the vapor phase. The hypothetical pressure is set equal to $P$ if it is $< P_{\rm sat}(T)$; otherwise we set $P = P_{\rm sat}$.}---condenses into dust particles. Interestingly, equation (\ref{eq:m_solid}) predicts that if the bubble shrinks isothermally, more dust condenses out. While such a trend seems sensible insofar as a decreasing volume increases the vapor density and by extension nucleation and condensation rates, the microphysics of dust growth is not actually captured by our model. As such, we do not know the condensate size distribution.

We assume for the purpose of calculating the bubble opacity to thermal radiation that all condensates are identical spheres of radius $s$ and internal density $\rho_{\rm solid} = 3\,{\rm g/cm}^3$. Further assuming that these spheres absorb and emit radiation with their geometric cross sections yields the optical depth of the bubble along its radius:
\begin{align}
\tau &=  \frac{9 M_{\rm solid}}{16\pi R^2 \rho_{\rm solid} s} \,.
\label{eq:tau}
\end{align} 

The bubble's temperature evolves with time $t$ according to the internal energy equation, whose right-hand side accounts for radiation, $PdV$ work, and latent heat release:
\begin{align} \label{eq:temp}
MC \frac{dT}{dt} = & - 4\pi R^2 \sigma_{\rm SB} (T^4-T_{\rm rad,bkg}^4) \, f(\tau) \nonumber \\
&- P\frac{dV}{dt} + L_{\rm vap}\frac{dM_{\rm solid}}{dt} 
\end{align}
where $C \simeq 3k/(\mu m_{\rm H}) = 8 \times 10^6\, {\rm erg/(g \,K)}$ is the bubble's specific heat (ignoring the order-unity difference between the specific heats of solids vs.~gas in the bubble), $\sigma_{\rm SB}$ is the Stefan-Boltzmann constant, and $L_{\rm vap} = 3 \times 10^{10} \,{\rm erg/g}$ is the latent heat of vaporization of rock.

The background radiation temperature, $T_{\rm rad,bkg}$, is the temperature to which the bubble would relax into radiative equilibrium in the absence of other effects. It is not necessarily the same as the background gas kinetic temperature $T_{\rm bkg}$, as the background can be optically thin, in which case $T_{\rm rad,bkg} < T_{\rm bkg}$. The bubble's emission likewise depends on its optical depth, through the dimensionless function $f(\tau)$. When the bubble is optically thin ($\tau \ll 1$), it is considered to have a uniform temperature $T$ and a uniform volume emissivity (power emitted per unit volume); accordingly, its luminosity scales as its volume $\propto R^2 \tau$. The same scalings apply for absorption of energy from the background radiation field in the optically thin limit. When the bubble is optically thick ($\tau \gg 1$), $T$ is interpreted as the internal ``core'' temperature of the bubble (where most of the bubble mass is assumed to reside), distinct from its photospheric temperature. When $T \gg T_{\rm rad,bkg}$, the photosphere is cooler than the core, allowing radiation to diffuse out at a rate proportional to $T^4/\tau$; in the opposite limit, the photosphere is hotter than the core, allowing radiation to diffuse in at a rate proportional to $T_{\rm rad,bkg}^4/\tau$ (note $\tau$ is a grey optical depth that is the same for outgoing and incoming radiation). We capture both optically thin and thick limits with the dimensionless 
function\footnote{The factor of $3\pi$ in $f(\tau)$ follows from the luminosity $\int j_\nu \, d\nu \times V$ of an optically thin sphere of constant volume emissivity $j_\nu = \alpha_\nu B_\nu$, where $\nu$ is the photon frequency, $B_\nu$ is the Planck function, and $\alpha_\nu = R/\tau$ is the grey absorption coefficient. The factor of 3/16 follows from the plane-parallel energy flux in radiative diffusion (assuming local thermodynamic equilibrium), $-(16/3)\sigma_{\rm SB} T^3 dT/d\tau$, and the approximation $dT/d\tau \sim -T/\tau$.}
\begin{align} \label{eq:f}
f(\tau) = \frac{\tau}{3\tau^2/16+3\pi} \,.
\end{align}
Our assumption that an optically thick bubble has a temperature gradient regulated by radiative diffusion across its entire radius ignores the possibility that latent heat release or turbulence might render the bubble more spatially isothermal, in which case the temperature difference $T-T_{\rm rad,bkg}$ may occur over an optical depth interval smaller than the bubble's full $\tau$, and thereby yield a higher radiative flux (either in or out). In this sense our picture is conservative because it minimizes radiative loss, staving off collapse.

The final equation governing the bubble describes how it expands or contracts depending on the pressure difference with the background:
\begin{align} 
M \frac{d^2 R}{dt^2} = 4\pi R^2 (P - P_{\rm bkg}) \label{eq:R} \,.
\end{align}
On the left-hand side we assume only the bubble mass accelerates and neglect the background mass that accelerates with it.\footnote{For oscillating air bubbles in water, the background water mass dominates the inertia (\citealt{minnaert_1933}).} The relevant background mass is located within distance $\sim$$R$ of the bubble surface, and as such makes only an order-unity contribution to the inertia when bubble and background densities are similar. As the bubble contracts and becomes more dense, the background inertia becomes less important. Note further that equation (\ref{eq:R}) assumes all of the bubble mass, including embedded solids, accelerates in response to gas pressure forces---condensates are assumed to be aerodynamically dragged in sync with bubble gas. This last assumption will be tested in section \ref{sec:drag} for various bubble parameters and particle sizes.


Equations (\ref{eq:temp}) and (\ref{eq:R}) are two coupled ordinary differential equations that we solve numerically (with Python's \texttt{odeint}). Note that the background gas kinetic temperature $T_{\rm bkg}$ does not formally enter into the equations solved.

\vspace{0.2in}

\subsection{Fiducial case}\label{sec:fiducial}
Input parameters evaluated at time $t=0$ of our fiducial model are as follows.

For the bubble: 
$T = 2500$ K, $P = P_{\rm sat}(T)$, $R = 1$ km, $\dot{R} = 0$, and $M = V \times \rho_{\rm sat}(T)$ so that $M_{\rm solid}=0$ (initially).

For the background: $T_{\rm rad,bkg} = 0$ and $P_{\rm bkg} = 0.99 P(t=0)$.

In words: a pure gas bubble starts stationary and slightly overpressured relative to the background, which has no radiation to heat the bubble. For reference, the bubble's initial pressure is $P(t=0)=9$ mbar, and its mass density $\rho_{\rm sat}(T) = 1.3 \times 10^{-6}$ g/cm$^3$ (about 3 orders of magnitude less than air at sea level). If the bubble mass $M$ were to compactify into a solid sphere of density $\rho_{\rm solid} = 3 \, {\rm g/cm}^3$, it would have a radius of 7.6 m. We set $s = 0.01$ cm, intermediate between the mm-cm sizes of the largest chondrules (e.g.~\citealt{simon_etal_2018}) and the $\mu$m sizes of matrix grains in chondritic meteorites (e.g.~\citealt{vaccaro_etal_2023}). The value of $s$ should not be taken too literally, as it is merely a proxy knob for the opacity, which in reality is determined by the (unmodeled) size distribution of condensed particles. The starting temperature of 2500 K matches the onset of condensation in the impact-vapor cloud models of Choksi et al.~(\citeyear{choksi_etal_2021}; their Fig.~1), and falls within the preferred range of temperatures in the ``metal-rich'' elemental condensation sequence for CB chondrites  (\citealt{campbell_etal_2002}, their Fig.~6). 

\begin{figure}
\includegraphics[width=0.95\columnwidth]{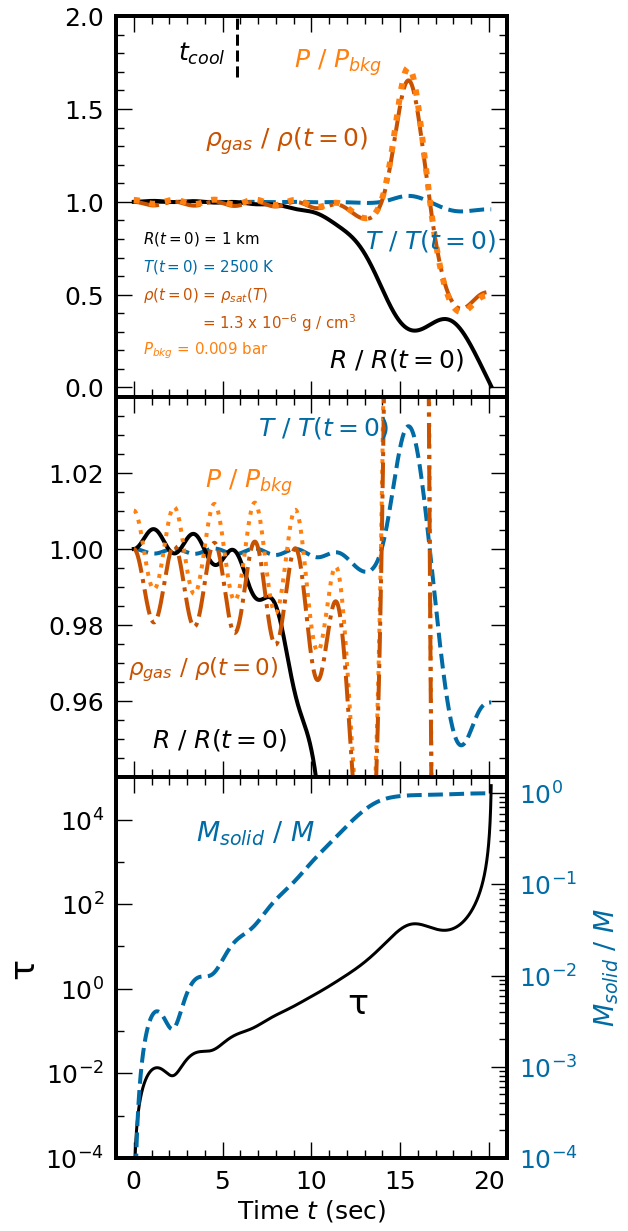}
\caption{Evolution of our fiducial bubble. The middle panel is the same as the top panel, zoomed in to show small variations in temperature, pressure, and density (gas density $\rho_{\rm gas}$ is distinct from the total density $\rho$ in solids+gas). The variations reflect fast acoustic oscillations and slower secular cooling by radiation. Initial conditions are listed in the top panel; additional inputs include the initial bubble velocity $\dot{R}(t=0)$ and the background radiation temperature $T_{\rm rad,bkg}$, both zero here. The bubble is seeded with a percent-level overpressure relative to the background (dotted orange line in middle panel) that causes it to first expand (solid black line), cool (dashed blue line), and condense (bottom panel showing dust optical depth $\tau$ and fractional mass in solids $M_{\rm solid}/M$). Radiation from this dust cools the bubble, and subsequent temperature oscillations trend ever so slightly downward. The corresponding secular decline in pressure causes the bubble to shrink. The smaller volume leads to still more condensation under saturated conditions (equation \ref{eq:m_solid}). When the bubble has nearly fully condensed ($t \gtrsim 14$ s), the order-unity pressure imbalance between bubble and  background results in a final implosion. The cooling time given by equation (\ref{eq:tcool}) is marked by a vertical dashed line.}
\label{fig:fiducial}
\end{figure}

Figure \ref{fig:fiducial} describes how such a bubble radiatively cools and eventually collapses. The initial overpressure sets the bubble oscillating. We verified that the oscillations, having period equal to the bubble sound-crossing time of
\begin{align} \label{eq:sound}
t_{\rm sound} \sim \frac{2R}{\sqrt{P/\rho_{\rm gas}}} \simeq 2 \left( \frac{R}{1 \, {\rm km}} \right) \, {\rm s}
\end{align}
last indefinitely when radiation is neglected (by setting $\tau = 0$ in equation \ref{eq:temp}). The oscillation is a simple acoustic or pressure mode (p-mode).

With every expansion, the bubble cools and condenses a fraction of its mass into dust. When radiation from this dust is accounted for, the bubble irreversibly leaks heat, and successive temperature minima decrease ever so slightly, producing more dust and cooling the bubble faster in a positive feedback. The temperature oscillations are small, less than a percent for the first 10 s of the evolution, and growing to only a few percent just prior to collapse. 

Pressure variations are more pronounced because the saturation vapor pressure depends exponentially on temperature. The upward trend in the amount of dust and the corresponding downward trend in the bubble's mean pressure causes a corresponding decline in its mean radius. In the penultimate oscillation ($t \simeq 14$--16 s), as the bubble shrinks to less than half its initial size, its internal gas pressure surges upward, but not enough to overcome the inertia of the infall. At this point, most of the bubble mass has condensed, not only from the decreasing temperature, but also from the decreasing volume (equation \ref{eq:m_solid}).

\begin{figure*}
\includegraphics[width=\textwidth]{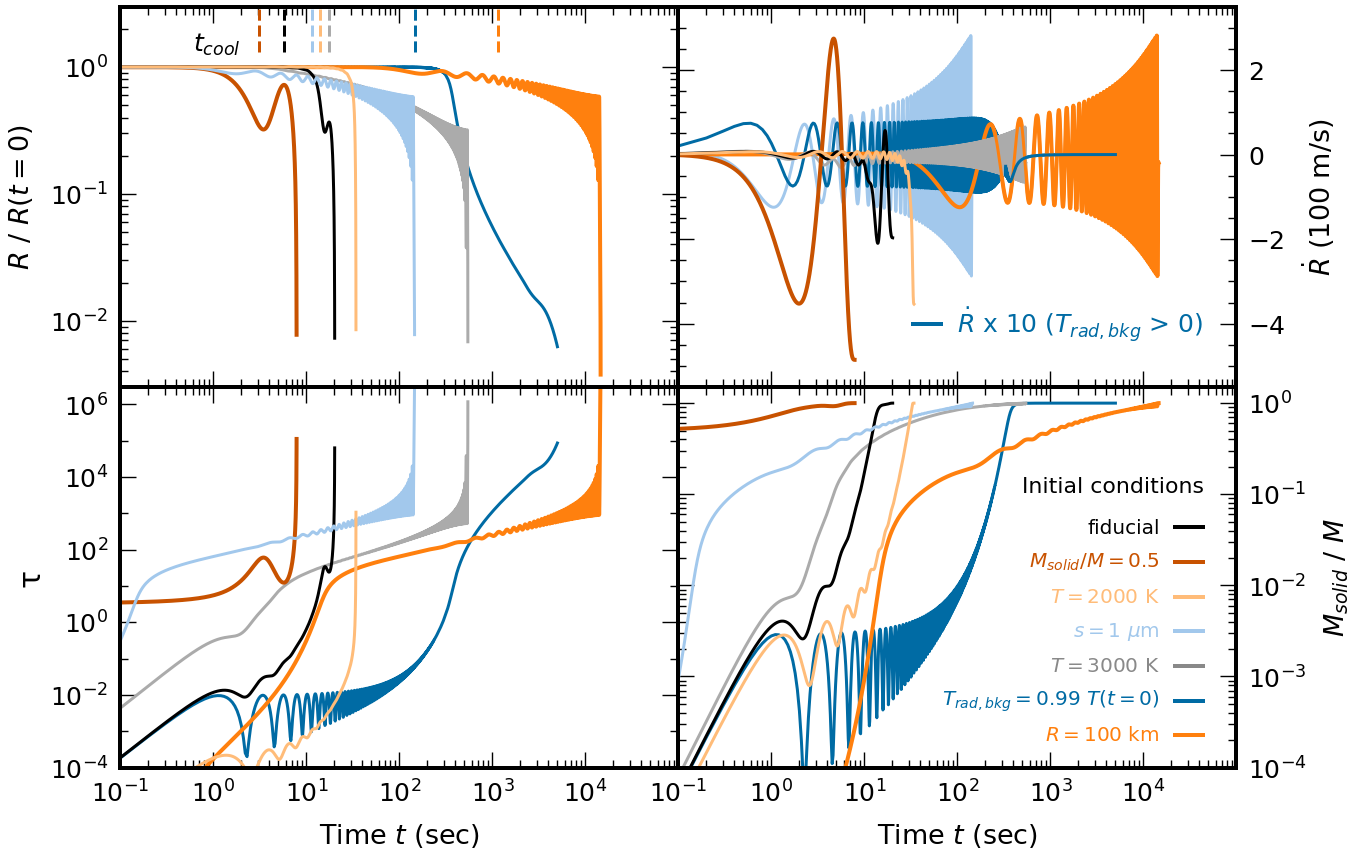}
\caption{Evolution of bubbles across parameter space. For each of the non-fiducial models, only one of the parameters as listed is changed, while other parameters are left at their fiducial values (namely $s = 100 \,\mu$m, $T_{\rm rad,bkg}=0$, $P_{\rm bkg} = 0.99P(t=0)$, and initial conditions $R = 1$ km, $\dot{R} = 0$, $T = 2500$ K, $M = \rho_{\rm sat}(T) V$, $P = P_{\rm sat}(T)$, and $M_{\rm solid}=0$). Estimated cooling times from equation (\ref{eq:tcool}), shown as vertical dashed lines in the top left panel, predict with varying accuracy the different collapse times (when $R \ll R(t=0)$). Discrepancies between $t_{\rm cool}$ and collapse times are due to evaluating $f(\tau)$ in  (\ref{eq:tcool}) at a single guessed value of $\tau$, when in reality $\tau$ varies with time. The longest collapse time, measured in hours, is associated with the $R = 100$ km bubble which contains the most mass to cool. The shortest collapse time of 8 seconds characterizes the bubble initialized with 50\% solid particles by mass; these seed solids provide an optical depth near unity at times before the final collapse, thereby maximizing $f(\tau)$ (equation \ref{eq:f}) and by extension the cooling luminosity. The bubble with $T_{\rm rad,bkg} > 0$ collapses slowest (for this model only, $\dot{R}$ in the top right panel is magnified by a factor of 10 for ease of viewing) and has its short-period pressure oscillations damped away at late times.}
\label{fig:param}
\end{figure*}

The bubble finally implodes at a near-sonic velocity from the order-unity difference in bubble vs.~background pressures. Radial flow speeds reach maxima of $\dot{R} \simeq -200$ m/s 
(see Figure \ref{fig:param}); for reference, the initial sound speed in the bubble is $\sim$$\sqrt{P/\rho_{\rm gas}} \simeq 800$ m/s. We follow the evolution until $R$ is $< 1\%$ of its initial value; formally, the total bubble density (in solids and gas) at this point is $\sim$1 g/cm$^3$,  practically that of a bulk solid. Despite the radius of the bubble shrinking by more than two orders of magnitude, the internal pressure $P$ and gas density $\rho_{\rm gas}$ vary by at most a factor of two from their initial values, and the temperature $T$ varies even less.

The collapse occurs over the time it takes the bubble to radiate away enough energy to condense. We define a characteristic cooling time by dividing the bubble's latent heat content by its radiative luminosity:
\begin{align} \label{eq:tcool}
t_{\rm cool} = \frac{ML_{\rm vap}}{\sigma_{\rm SB} (T^4-T_{\rm rad,bkg}^4) 4\pi R^2 f(\tau)} \,.
\end{align}
The numerator ignores the contribution in heat content from changes in temperature since the bubble collapses nearly isothermally. Equation (\ref{eq:tcool}) can be evaluated using initial conditions, setting $\tau$ to what it would be if $M_{\rm solid} = 0.3 \, M$ and $R = R(t=0)$ (the factor of 0.3 is an arbitrary order-unity constant chosen to approximately match numerical results). So defined, $\tau \simeq 1$ and $t_{\rm cool} \simeq 6$ s (see vertical dashed line in the top panel of Fig.~\ref{fig:fiducial}). This calculation underestimates the actual collapse time of $\sim$20 s because in reality the bubble spends the first half of its evolution optically thin and radiating relatively inefficiently.

\subsection{Parameter study}\label{sec:param}
The collapse of the bubble as described in the previous section occurs irrespective of many parameter choices. In particular, the initial pressure difference between bubble and background can vary by orders of magnitude, and be negative or positive or zero, and the bubble will still collapse, as we have verified by direct calculation (data not shown). Assuming the background can serve as an energy sink ($T > T_{\rm rad,bkg}$), all that is needed for collapse is dust to secularly cool the bubble. Despite formally having no dust to begin with, our fiducial bubble becomes dusty because it is initially overpressured relative to the background, and so expands, cools, and condenses. Other combinations of initial pressure difference $P-P_{\rm bkg}$ (including zero) and initial velocity perturbation $\dot{R}$ (including zero) lead to similar runaway outcomes, where cooling, condensation, and contraction feed off one another in a positive feedback. Simply seeding the bubble with dust ($M_{\rm solid} > 0$ at $t=0$) also works.

Figure \ref{fig:param} shows how the collapse of the bubble varies quantitatively with some input parameters. We change one of the following five parameters at a time, keeping others at their fiducial values:
\begin{enumerate}
\item {\em Seed solid mass}: $M_{\rm solid}(t=0)/M = 0.5$ (red curves in Fig.~\ref{fig:param}). Compared to the fiducial case with no seed solids (black curves), the bubble with seed solids is optically thicker, radiates faster, and collapses on a timescale shorter than that of the fiducial model by a factor of $\sim$2, and reasonably predicted by $t_{\rm cool}$ (3 s; vertical red dashed line in top left panel). Radial flow speeds are correspondingly faster, with a maximum contraction velocity of 500 m/s.
\item {\em Particle size controlling opacity}: $s = 1 \, \mu$m (light blue curves). Smaller particles have higher radiative cross-sections per unit mass and render the bubble optically thicker. At first the bubble cools and shrinks faster than in the fiducial model (black curves, $s = 100$ $\mu$m), but later becomes so much more opaque that it cools slower 
and ultimately takes longer to collapse (150 s vs.~20 s).
\item {\em Background radiative heating}: $T_{\rm rad,bkg} = 2475 \, {\rm K} = 0.99 \,T(t=0)$ (dark blue curves). Cooling is slowed relative to the fiducial model (black curves, $T_{\rm rad,bkg}=0$), with collapse taking $\sim$1 hour. Our metric $t_{\rm cool}$, though lengthened by $T_{\rm rad,bkg}$ according to the denominator of equation (\ref{eq:tcool}), still underestimates the actual collapse time because the formula does not consider how background heating limits dust production and keeps the optical depth low for some time (bottom left panel). Background heating buffers the bubble against the large-amplitude oscillations in radius seen in other models. Radial speeds are highly sub-sonic (and are magnified for easier viewing in Fig.~\ref{fig:param} by a factor of 10, for this model only), varying between $\pm$ 8 m/s during the initial quasi-adiabatic acoustic-oscillation phase, and slowing from -6 m/s to -0.4 cm/s during the protracted final collapse. As this experiment demonstrates, the bubble can leak energy to the environment and collapse even when $T_{\rm rad,bkg} < T$ by as little as -1\%. Various experiments with $T_{\rm rad,bkg} > T(t=0)$, even by +1\%, do not lead to collapse (data not shown).
\item {\em Initial bubble size}: $R(t=0) = 100$ km (heavy orange curves). Compared to our fiducial 1-km bubble (black curves), the 100-km bubble has a 100$\times$ longer acoustic period (200 s vs.~2 s because of the longer distance crossed at the same sound speed). Collapse takes several hours, about 3 orders of magnitude longer than the fiducial case, because the larger bubble has more mass to cool and its radiative luminosity does not scale in linear proportion to the mass when the bubble becomes opaque. Equation (\ref{eq:tcool}) predicts $t_{\rm cool} \propto M/(R^2 f(\tau))$, which in the optically thick limit scales as $M/(R^2/\tau) \propto R^2$ (and in the optically thin limit scales as $M/(R^2\tau) \propto M^0$). 
\item {\em Initial bubble temperature}: $T(t=0) = 3000$ K (grey curves) and $2000$ K (light orange curves). Higher temperatures imply higher densities from the saturation pressure curve (equation \ref{eq:psat}) and the ideal gas law. At 3000 K, the bubble is 17$\times$ denser than our fiducial 2500 K bubble (black curves); the hotter bubble, if compactified to solid density, would yield a planetesimal 19 m in radius (vs.~7.6 m for the fiducial bubble).  
The correspondingly larger mass takes longer to cool and collapse, about
10 minutes, with radial speeds varying between $\pm$ 60 m/s.
Compared to the fiducial model, the cooler 2000 K bubble is 43$\times$ less massive (equivalent planetesimal radius 2.2 m), but also takes longer to collapse (34 s vs. 20 s) because it generates comparatively little dust (bottom left panel) and therefore cools less efficiently.
\end{enumerate}

In 5 out of 6 of our non-fiducial, parameter space experiments, the bubble pressure and gas density deviate by less than a factor of 10 relative to their initial values, and for temperature much less, despite reductions in bubble volume by more than a factor of $10^6$. The smallest fractional changes are seen in the background heating experiment ($T_{\rm rad,bkg} = 0.99\, T(t=0)$), where $\Delta T/ T \lesssim 10^{-3}$ (thereby maintaining the inequality $T > T_{\rm rad,bkg}$ throughout the evolution) and $\Delta P / P \sim \Delta \rho_{\rm gas} / \rho_{\rm gas} \lesssim 2\%$. In the large bubble (initial $R$ = 100 km), hotter temperature (initial $T = 3000$ K), seed solid (initial $M_{\rm solid}/M = 0.5$), and small particle size ($s = 1$ $\mu$m) experiments, temperatures deviate by 2-14\% relative to starting conditions, and pressures and gas densities range from 0.1$\times$ to 6$\times$ initial values. The largest variations are seen for the cold bubble (initial $T = 2000$ K), whose final temperature is $T \simeq 1550$ K, and whose final (saturation) pressure is accordingly a factor of 180 smaller than the initial pressure.

\begin{figure*}
\includegraphics[width=\textwidth]{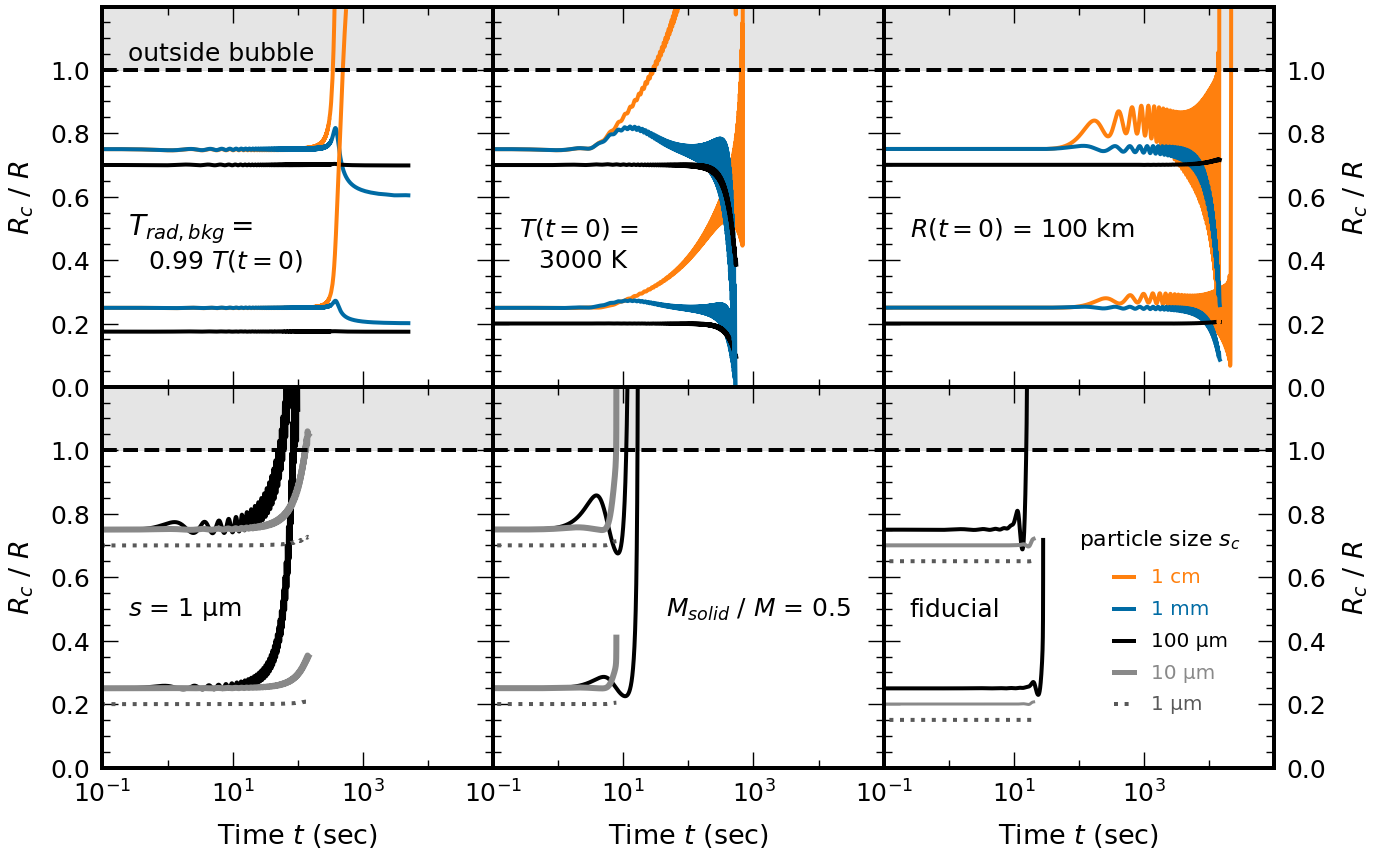}
\caption{Trajectories of solid particles of varying sizes $s_{\rm c}$ (distinct from $s$, the size of the particles controlling opacity) within six of the model bubbles shown in Fig.~\ref{fig:param} (omitting the cold, initial $T = 2000$ K model). The particles' radial locations $R_{\rm c}$ are normalized to the bubble radius $R$. Particles are initially placed at either $R_{\rm c}/R = 0.25$ or 0.75; to more easily distinguish between curves, some have been offset in the ordinate, or scaled in time by order-unity factors.
The largest particles in each panel fail to be entrained by bubble gas and are stranded at $R_{\rm c}/R > 1$ while the bubble collapses. Smaller particles stay inside and in some cases are propelled to the center somewhat faster than gas.
}
\label{fig:peb}
\end{figure*}

\vspace{0.3in}
\subsection{Particle drag and relative velocities}\label{sec:drag}
Here we evaluate how well bubble gas entrains solid particles of various sizes. We consider a solid particle of radius $s_{\rm c}$ and mass $m_{\rm c} = 4\pi\rho_{\rm solid}s_{\rm c}^3/3$ located at radial distance $R_{\rm c} < R$ from the bubble center. The particle accelerates by gas drag according to
\begin{align} \label{eq:drag}
m_{\rm c} \frac{d^2 R_{\rm c}}{dt^2} & = \frac{4\pi}{3} \rho_{\rm gas} \left( u_{\rm gas} - \frac{dR_{\rm c}}{dt} \right) \sqrt{\frac{8kT}{\pi \mu m_{\rm H}}} \, s_{\rm c}^2 \, g(s_{\rm c}/\lambda_{\rm mfp}) \\
g & = \frac{1}{1+0.63s_{\rm c}/\lambda_{\rm mfp}} \label{eq:drag1}
\end{align}
where $\lambda_{\rm mfp} = \mu m_{\rm H}/(\rho_{\rm gas} \sigma)$ is the collisional mean free path in gas with molecular cross-section $\sigma \simeq \pi \times 10^{-15}$ cm$^2$. The dimensionless function $g$ is constructed to smoothly switch from a drag force that scales as $s_{\rm c}^2$ when $s_{\rm c} \ll \lambda_{\rm mfp}$ (free molecular drag; \citealt{epstein_1924}), to a force that scales as $s_{\rm c}\lambda_{\rm mfp}$ when $s_{\rm c} \gg \lambda_{\rm mfp}$ (Stokes drag).\footnote{Numerical coefficients in (\ref{eq:drag}) and (\ref{eq:drag1}) match coefficients derived in the literature. In the free molecular drag regime, we use the force law that assumes ``specular reflection'' of molecules off a sphere (\citealt{epstein_1924}). In the Stokes regime, the drag force on a sphere moving at speed $u$ relative to gas is $6\pi \rho_{\rm gas} \nu_{\rm visc} u s_{\rm c}$, with kinematic viscosity $\nu_{\rm visc} = 0.35 \lambda_{\rm mfp} u_{\rm thermal}$ and $u_{\rm thermal} = \sqrt{8kT/(\pi \mu m_{\rm H})}$ (\citealt{chapman_cowling_1970}).}
The gas velocity $u_{\rm gas}$ at the particle's position is scaled to $\dot{R}$, the velocity at the bubble boundary:
\begin{align} \label{eq:hubble}
u_{\rm gas} = \dot{R} \times R_{\rm c}/R \,.
\end{align}
Thus $u_{\rm gas} = 0$ when $R_{\rm c}=0$ at the bubble center, and grows linearly up to $\dot{R}$ at $R_{\rm c} = R$ (a linear ``Hubble law'').

Equation (\ref{eq:drag}) is solved numerically for $R_{\rm c}(t)$, with $R(t)$ and $\dot{R}(t)$ obtained separately from the equations in section \ref{sec:basic}. The solid particle is thus treated as a kind of test particle in the gas; the momentum backreaction of solids on gas is ignored. This neglect means our solutions for $R_{\rm c}(t)$ at late times, when solids outweigh gas, cannot be trusted. Nevertheless the calculations provide a starting point for discussion, and should still identify reasonably accurately the largest particles that can compactify with the bubble. It is hard to see how accounting for drag backreaction would change the sizes above which particles would decouple from the gas.

Figure \ref{fig:peb} displays the evolution of particle positions relative to the bubble boundary, $R_{\rm c}/R$, for particles of various sizes, in our fiducial model (section \ref{sec:fiducial}) plus five of our experiments exploring parameter space (section \ref{sec:param}, omitting the cold bubble experiment). For each model at $t=0$ we place initially stationary particles ($\dot{R}_{\rm c} = 0$) at $R_{\rm c}/R = 0.25$ and 0.75. In plotting $R_{\rm c}/R$ vs.~$t$, we offset some curves on the ordinate or stretch them on the abscissa for clarity.

The smallest particles, having sizes from 1 to 100 $\mu$m depending on the model, are so tightly coupled to gas that they maintain their relative positions in the bubble; their $R_{\rm c}/R$ curves are nearly flat, so that when the bubble collapses, 
they collapse along with it. The largest particles, having sizes of 10 $\mu$m to 1 cm depending on model parameters, are left outside the bubble when it implodes (we stop integrating particle trajectories when $R_{\rm c}/R > 1$). Intermediate size particles stay within the bubble and change their positions within it, moving radially outward or inward, not quite in sync with bubble gas.

Particles are least coupled to those bubbles that collapse fastest, as can be seen in Fig.~\ref{fig:peb} by comparing the bottom row of faster-collapsing models to the top row of slower-collapsing models. Across the parameter space surveyed, cm-sized particles and larger do not collapse. The drag regime is largely controlled by the gas temperature, which sets the gas density under saturation conditions. For $T \leq 2500$ K, gas mean free paths $\lambda_{\rm mfp}$ exceed 100 $\mu$m, and smaller sized particles experience free molecular drag. In this regime the momentum stopping time is 
\begin{align}
t_{\rm stop} \equiv \frac{m_{\rm c}u_{\rm rel}}{\frac{4\pi}{3} \rho_{\rm gas} u_{\rm rel} \sqrt{\frac{8kT}{\pi \mu m_{\rm H}}}s_{\rm c}^2} = \frac{\rho_{\rm solid} s_{\rm c}}{\rho_{\rm gas} \sqrt{\frac{8kT}{\pi \mu m_{\rm H}}}}
\end{align}
which evaluates to $\leq 0.2$ s for $s_{\rm c} \leq 100$ $\mu$m, $T = 2500$ K, and $\rho_{\rm gas} = \rho_{\rm sat}(T)$. These stopping times are shorter than our bubble acoustic oscillation periods, which are of order $t_{\rm sound} \sim 2$--200 s (equation \ref{eq:sound}), ensuring entrainment. Larger particles with $s_{\rm c} > \lambda_{\rm mfp}$ will have longer stopping times, not only because of their larger inertia but also because they experience Stokes drag which is less efficient than free molecular drag by a factor of $s_{\rm c}/\lambda_{\rm mfp}$. For model bubbles that take hours to collapse, the cut-off size for retaining particles appears to be a millimeter (top row of Fig.~\ref{fig:peb}).

Particles within a bubble will have a dispersion of sizes, and accordingly different velocities relative to the gas, and different velocities relative to one another. Particle-particle velocities $u_{\rm rel}$ are limited by $|\dot{R}|$, the maximum gas velocity at the bubble boundary (equation \ref{eq:hubble}); a hard upper limit on $u_{\rm rel}$ would be $2\dot{R}$, if we imagine one particle traveling inward at $-|\dot{R}|$ and a second traveling outward at $+|\dot{R}|$ (accelerated to this velocity during a prior expansion phase of the bubble, and somehow left coasting). For most particle-particle encounters, $u_{\rm rel} < |\dot{R}|$, with smaller particles, and smaller differences in sizes between particles, reducing $u_{\rm rel}$.

For cavitating bubbles to collect chondrules into chondrite parent bodies, relative particle velocities cannot be too large, lest chondrules be destroyed on impact. The sedimentary laminations of the CB/CH chondrite Isheyevo point to gentle, layer-by-layer accretion of size and mineral-sorted material (\citealt{garvie_etal_2017}). \citet{wurm_etal_2005} and \citet{teiser_etal_2009} require $u_{\rm rel} \lesssim 13$--25 m/s for mm-sized aggregates to not break on impact. The aggregates in their experiments are porous and composed of micron-sized monomers. Chondrules and metal nodules are individually competent and should tolerate higher impact velocities; on the other hand, they may splash when molten. \citet{hood_ciesla_2001} cite upper limits on $u_{\rm rel}$ of $\sim$100 m/s for solid chondrules, and $\sim$1 m/s for liquid ones. \citet{garvie_etal_2017} roughly estimate $u_{\rm rel} \lesssim$ 30--200 m/s based on the material strengths of carbonaceous chondrites. These variously estimated maximum velocities should be compared against our theoretical values of $\dot{R}$ (read $\max u_{\rm rel}$) displayed in Fig.~\ref{fig:param}. Of particular interest for chondrite assembly are the slowest velocities $|\dot{R}| <$ 8 m/s for our model bubble heated by background radiation.

\begin{figure*}
\includegraphics[width=\textwidth]{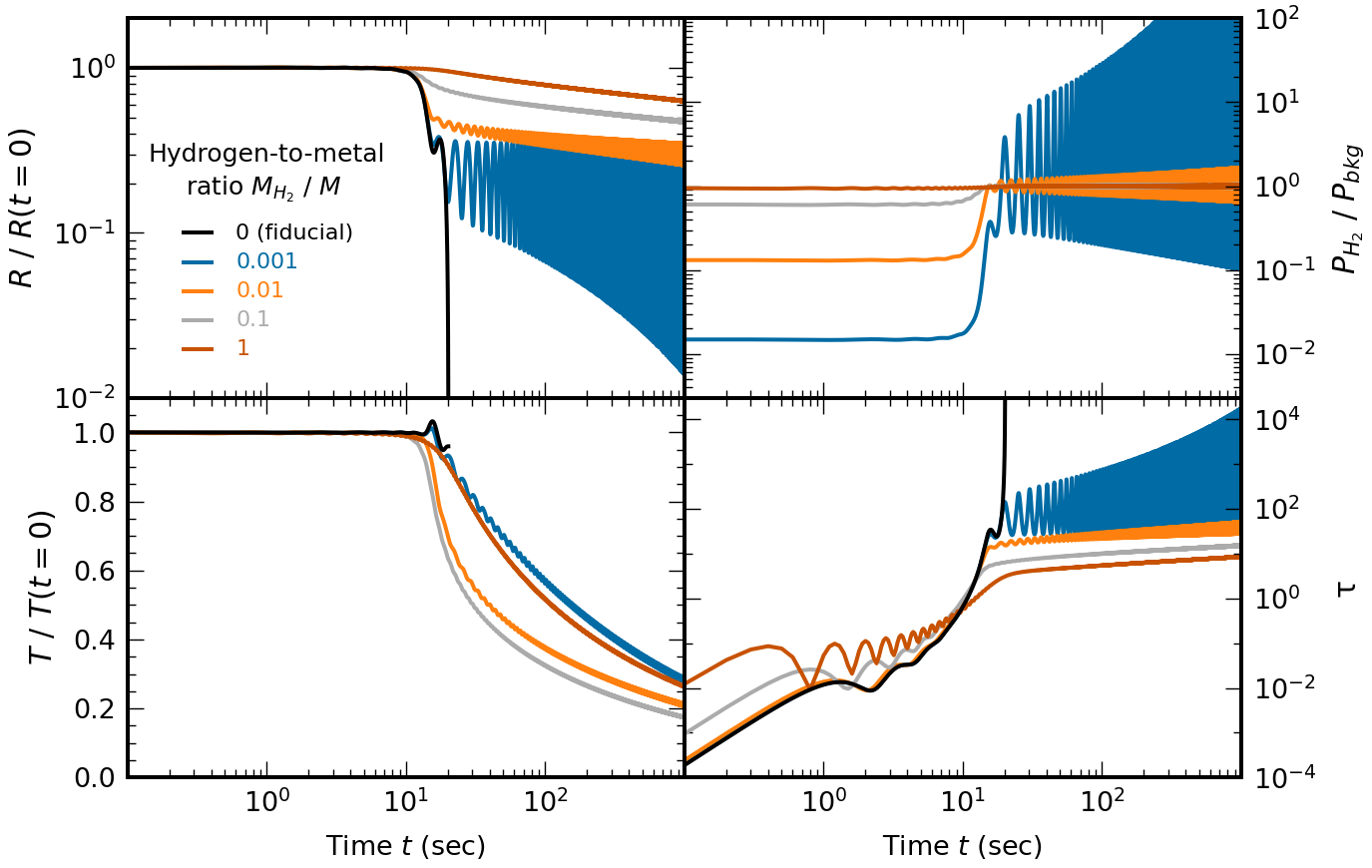}
\caption{Adding hydrogen, a non-condensible gas, to the bubble maintains a minimum pressure and prevents wholesale collapse. Aside from the H$_2$ fraction, all other parameters of the models shown here are the same as in the fiducial model. Even small hydrogen mass fractions can effectively stabilize the bubble. The initial hydrogen partial pressure $P_{\rm H_2}$ is larger than the initial saturation vapor pressure of silicates by the ratio of mean molecular weights $\mu/\mu_{\rm H_2} \simeq 15$. When $M_{\rm H_2}/M = 0.01$, the hydrogen partial pressure $P_{\rm H_2}$ sits just a factor of 10 below the background pressure $P_{\rm bkg}$ (assumed to be nearly equal to the bubble's initial total pressure). This factor of 10 is recovered, and pressure equilibrium with the background achieved, once the bubble shrinks by a factor of $\sim$2 from the loss of silicate vapor pressure (top panels). At this point all of the silicate is condensed; the dust particles, now residing in inert H$_2$, are free to radiatively cool to the background radiation temperature (here zero), further shrinking the bubble but more slowly compared to earlier. From $dT/dt \propto R^2 T^4 / \tau$ (for optically thick bubbles; see bottom right panel), $\tau \propto R$, and $T/V =$ constant (constant pressure maintained by the background), it follows that $T \propto t^{-3/10}$ and $R \propto t^{-1/10}$. We have verified that secular cooling and shrinkage halts once $T$ reaches $T_{\rm rad,bkg} > 0$ (data not shown).
}
\label{fig:gas}
\end{figure*}

\section{Bubbles of Condensible Vapor Mixed with Hydrogen} \label{sec:gas}
We now add hydrogen, a non-condensing gas, to the silicate vapor in the bubble. In equations (\ref{eq:temp}) and (\ref{eq:R}) we make the following replacements:
\begin{align}
MC & \rightarrow MC + M_{\rm H_2}C_{\rm H_2} \\
P & \rightarrow P + P_{\rm H_2} \\
M & \rightarrow M + M_{\rm H_2}
\end{align}
where variables subscripted ${\rm H_2}$ refer to molecular hydrogen and those not so subscripted refer to the condensible silicates. The specific heat $C_{\rm H_2} = (7/2)k / (\mu_{\rm H_2} m_{\rm H}) = 1.5\times 10^8 \, {\rm erg}/({\rm g} \, {\rm K})$ where $\mu_{\rm H_2} = 2$, and $P_{\rm H_2} = (M_{\rm H_2}/V) kT / (\mu_{\rm H_2} m_{\rm H})$. 

Figure \ref{fig:gas} shows how the evolution of our fiducial bubble changes with increasing hydrogen-to-metal mass fraction $M_{\rm H_2}/M$. A tiny addition of hydrogen can substantially slow the collapse. The first $\sim$10-20 s of the evolution is largely independent of $M_{\rm H_2}$, with nearly all the metals condensing out before 20 s. But whereas condensation removes all of the bubble's pressure support when $M_{\rm H_2} = 0$, the presence of ${\rm H_2}$ sets a floor on the pressure. Pound for pound at a given temperature, ${\rm H_2}$ supplies more pressure than silicate vapor by a factor of $\mu/\mu_{\rm H_2} \sim 15$; the reduction in bubble volume immediately following metal condensation quickly amplifies the ${\rm H_2}$ pressure up to the background pressure (top right panel of Fig.~\ref{fig:gas}). Thereafter the bubble continues to shrink from radiative cooling, albeit more slowly, remaining on average in pressure equilibrium with the background while becoming colder and denser. From $dT/dt \propto R^2 T^4 / \tau$ (for optically thick bubbles), 
$\tau \propto R$, and $T/V =$ constant (constant pressure maintained by the background), we derive $T \propto t^{-3/10}$ and $R \propto t^{-1/10}$, as confirmed in Fig.~\ref{fig:gas}.

Experiments that simultaneously vary $M_{\rm H_2}/M$ with solid seed mass or initial bubble temperature do not yield behaviors beyond those reported above. When both $M_{\rm H_2}$ and the background radiation temperature $T_{\rm rad,bkg} < T(t=0)$ are non-zero, the bubble shrinks and cools until $T = T_{\rm rad,bkg}$, at which point the bubble stabilizes.

\begin{figure*}
\includegraphics[width=\textwidth]{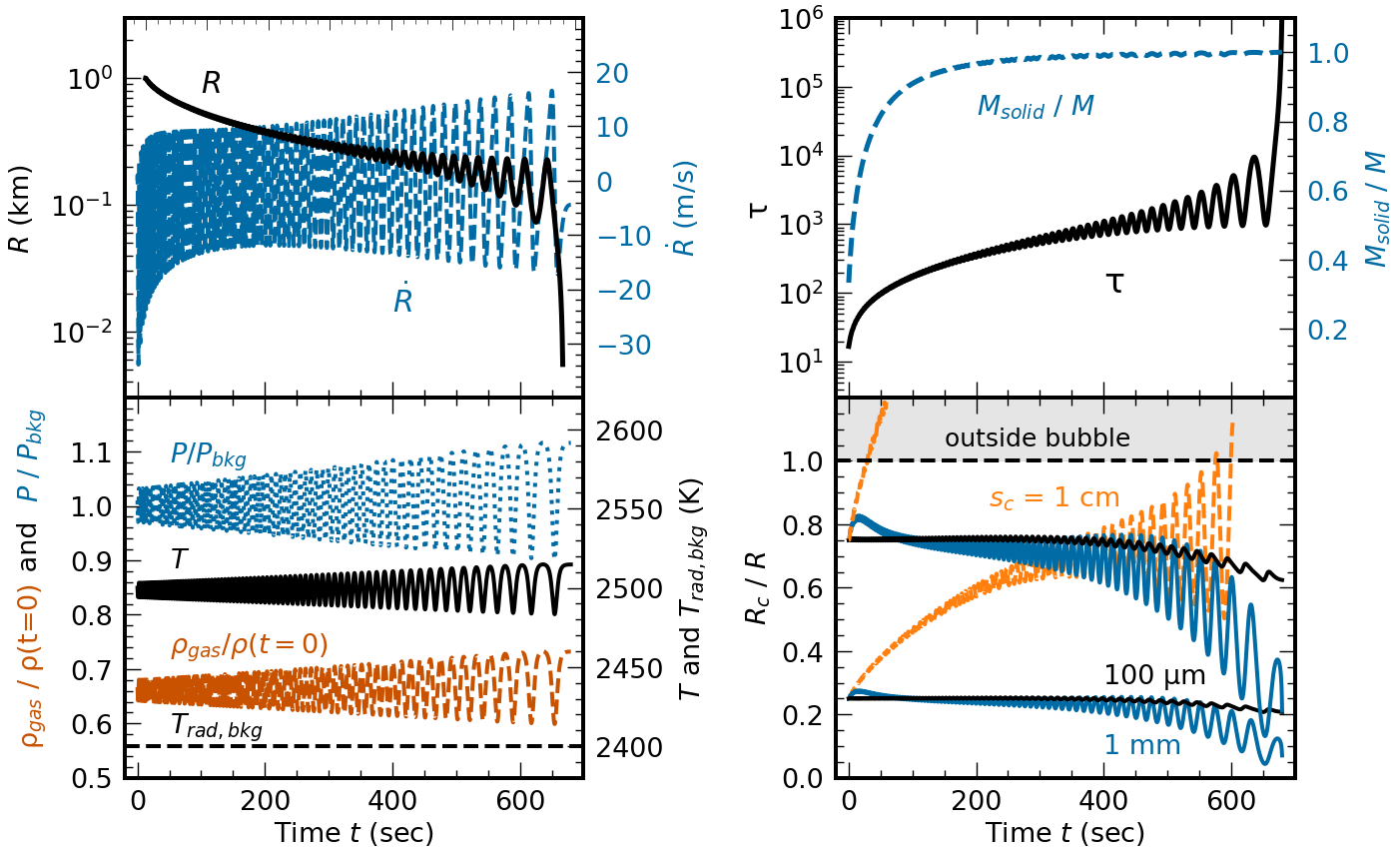}
\caption{A cavitating bubble of silicate/metal vapor and dust (no H$_2$ anywhere), set against the backdrop of an impact plume generated by a vaporizing collision between asteroids (e.g.~\citealt{choksi_etal_2021}), could form a chondritic parent body. Here an $R = 1$ km sized bubble starts optically thick at $t=0$ (top left and right panels), with $M_{\rm solid}/M = 1/3$ of its mass in $s = 10 \,\mu$m sized dust grains, and the remaining 2/3 in saturated silicate vapor. The bubble resides in a radiation bath whose temperature $T_{\rm rad,bkg}$ is a few percent lower than the bubble's (bottom left panel), allowing the bubble to radiate energy to the background and ultimately collapse after $t \simeq 10$ minutes. Collapse speeds vary up to $\sim$20 m/s (top left panel), fast enough to overcome background flows (estimated to have a magnitude of $\lesssim 0.3$ m/s on $R \lesssim 1$ km scales; see main text) trying to shear the bubble apart. Embedded particles chondrule-sized or smaller, with sizes $s_{\rm c} \leq 1$ mm, compactify with the bubble (bottom right panel), while cm-sized and larger particles are too heavy to be dragged by gas and are left behind. Particle relative velocities are expected to be less than $\max |\dot{R}| \simeq 20$ m/s, slow enough that colliding particles may not shatter \citep{hood_ciesla_2001,wurm_etal_2005,teiser_etal_2009,garvie_etal_2017}. Compactified to solid density ($\rho_{\rm solid} = 3$ g/cm$^3$), the bubble would produce a planetesimal of radius 9 m.
}
\label{fig:the_one}
\end{figure*}

\section{Summary and Discussion}\label{sec:sum}
A condensing gas loses pressure as vapor converts to liquid or solid particles. We have raised the possibility that condensation proceeds inhomogeneously---that pockets or ``bubbles'' of gas that begin condensing relative to their surroundings will accelerate their condensation and collapse under background pressure. We have calculated how a bubble shrinks as its condensates (``dust'') radiate energy to the background, under the drastic simplifying assumption that background properties (pressure and radiation field) stay constant. For the background to serve as an energy sink, its radiation temperature must be less than the bubble's. 


Gas condenses over the time it takes radiation to carry away latent heat. Bubbles having sizes of 1--100 km, saturated with silicate vapor at temperatures of $\sim$2500 K, can condense over seconds to hours, depending on parameters. The temperature stays nearly constant during this time. Once condensation completes, the bubble implodes, at near-sonic speed in the absence of background heating (similar to cavitating water vapor bubbles on Earth), and at subsonic speeds (m/s or below) in the presence of background irradiation.

Adding a non-condensible gas like ${\rm H_2}$ to the bubble sets a floor on the pressure, halting the collapse when the bubble, containing only inert dust and ${\rm H_2}$, attains radiative and pressure equilibrium with the background. The resultant clumps may have astrophysical application. Perhaps the clumpy solids revealed by transits of pre-main-sequence stars (e.g.~\citealt{bodman_quillen_2016}), or white dwarfs (e.g.~\citealt{vanderburg_etal_2015}; \citealt{powell_etal_2021}), arise from a stalled form of cavitation, or a thermal instability that generates a multi-phase medium (\citealt{field_1965}). Orbital shear sets an upper limit of about an orbital period on the lifetimes of non-self-gravitating bubbles/clumps.


The obvious shortcoming of our work is that we have solved for the evolution of a bubble in a background medium while not accounting for how the background medium may also change. A linear stability analysis of the background, with an imposed cooling function to drive the medium to condense over time, may be a good next step. The question is whether small perturbations to the background undergo runaway condensation and collapse, and on what spatial and time scales. If the medium is optically thick, radiation can be treated locally in a diffusive approximation; but if optically thin, the problem will be global, with the radiation field at one point dependent on radiation emitted from elsewhere. The need for collapsing bubbles to lose energy to the background suggests that while the bubbles may be optically thick, the background may need to be optically thin, or at least thinner than the bubbles. Other issues requiring investigation include turbulence; instabilities at the bubble boundary (e.g.~Rayleigh-Taylor; Richtmyer-Meshkov); the drag backreaction of solids on gas; and the evolution of the condensate size distribution. 

\vspace{0.3in}
\subsection{Pros and cons of cavitation as a means of assembling chondrites}
Might cavitation be a way to accrete chondrules and other condensates into chondrite parent bodies---or at least create overdensities sufficiently strong for self-gravity to take over (e.g.~\citealt{alexander_etal_2008})? Cooling and condensation, and the compression resulting therefrom, are generic processes which should occur in some form whatever the nature of the heating event that melted chondrules (asteroid collisions, nebular shocks, stellar flares, lightning, ...). That cavitation can happen quickly, in the immediate aftermath of heating, on timescales possibly shorter than a day, helps explain the large volume filling fractions of chondrules in meteorites. Cavitation promises to collect melt droplets together as soon as the flash-heated vapor starts to cool and condense; there would be little intervening time for chondrules to mix and be overly diluted by other kinds of protoplanetary disk solids.

Cavitation of condensing silicate vapor is also consistent with the size distribution of chondrite constituents, ranging from the $\mu$m sizes of matrix grains (e.g.~\citealt{vaccaro_etal_2023}) to the 0.1-1 mm sizes of chondrules (e.g.~\citealt{simon_etal_2018}). We have shown over a wide range of parameter space that particles having such sizes can be entrained by collapsing bubbles, whereas cm-sized and larger particles may fail to be dragged along and are  stranded outside. The particle size dividing coupled and uncoupled particles is around 1 mm, which follows naturally from the characteristic temperatures (a few thousand deg K) and pressures (measured in mbar) of saturated silicate vapor, and the corresponding aerodynamic stopping times. Background radiative heating can slow bubble collapse speeds to below $\sim$10 m/s and prevent destruction of particles as they are agglomerated.

That bubbles can collapse unimpeded when they are not contaminated by a non-condensible gas like ${\rm H_2}$ appears to favor chondrule formation by asteroid collisions, as opposed to a nebular scenario involving H$_2$. The case for a collisional origin has been made most convincingly for the CB/CH subclass of chondrules, whose elemental abundances point to condensation in a practically pure silicate/metallic vapor (e.g.~\citealt{campbell_etal_2002}; \citealt{krot_etal_2007}). Silicate chondrules and Fe/Ni metal nodules have exceptionally high volume filling fractions in CB/CH chondrites ($>95\%$; e.g.~\citealt{jacquet_2022}), underscoring the need for a rapid and efficient accretion mechanism, which cavitation offers. If there were any ambient nebular hydrogen at the time of the collision---and see \citet{krot_etal_2005} and \citet{wolfer_etal_2023} for radiogenic evidence that little or no H$_2$ remained at the time of CB chondrite formation---the H$_2$ would be evacuated to the periphery of an expanding cloud of silicates and metals (\citealt{choksi_etal_2021}).

\citet{stewart_lpsc6,stewart_lpsc7}, \citet{stewart_lpsc3}, and \citet{carter_stewart_2020} highlight how nebular H$_2$ confines the impact plume from an asteroid collision and forces its re-collapse (see also criticism of this scenario by \citealt{choksi_etal_2021}, as reviewed in our section \ref{sec:intro}). Our emphasis here is different. We have imagined instead an H$_2$-free scenario, where collapse occurs locally, on small scales, within a plume that consists wholly of condensibles and is seeded with small thermal fluctuations. In an expanding vapor plume, collapse can only proceed on small scales, as shear from the background expansion tears apart material on large scales. In a plume of length scale $\mathcal{L}$ whose boundary expands outward at velocity $\mathcal{\dot{L}}$, velocity differences between points separated by $R < \mathcal{L}$ are of order $\Delta {\mathcal{\dot{L}}} \sim \mathcal{\dot{L}} \times R/\mathcal{L}$ (free expansion). For $\mathcal{\dot{L}}$ to be less than a bubble collapse speed of, say, 30 m/s, in a plume having $\mathcal{L} \sim 10000$ km and $\mathcal{\dot{L}} \sim$ 3 km/s (e.g.~\citealt{choksi_etal_2021}), cavitation would be restricted to bubble sizes $R \lesssim 100$ km. 

In Figure \ref{fig:the_one} we offer a summary bubble evolution that addresses the above constraints on chondrite formation---but also points to problems. The 1-km bubble starts optically thick, with solid particles 10 $\mu$m in size comprising 1/3 of the total mass, and the remaining 2/3 in saturated silicate vapor. The bubble maintains a temperature of 2500 K, the temperature marking the onset of condensation in the expanding impact-plume models of \citet{choksi_etal_2021}, and favored by the metal-rich condensation sequences of \citet{campbell_etal_2002} for CB chondrites.  The bubble collapses within about 10 minutes, shorter than the backgroud plume evolution timescale measured in hours \citep{choksi_etal_2021}. Gas velocities within the bubble peak at $\sim$20 m/s, fast enough to defeat the $\sim$0.3 m/s $(10^4 \, {\rm km}/\mathcal{L}) [\mathcal{\dot{L}}/(3 \,{\rm km/s})]$ shear from the impact plume stretching the bubble apart. Embedded solid particles that are 1 mm in size or smaller follow the bubble as it shrinks. When particles collide with one another, they do so at speeds $<$ 20 m/s (and probably considerably less than that depending on the particle size distribution), slow enough to avoid shattering.

\vspace{0.05in}
We identify a couple problems with this scenario:
\begin{enumerate}
\item For the bubble to shrink, the gas within it should have a lower time-averaged pressure than the background gas, which means the bubble gas should be colder (on average) than the background gas. At the same time, for the bubble to leak radiation to the background and cool, the emitting particles in the bubble should be hotter than the background radiation. Satisfying both conditions (which the model in Fig.~\ref{fig:the_one} does by fiat, by setting the free parameter $T_{\rm rad,bkg} < T$) would seem to require the background to be optically thin, or at least thinner than the bubble. This would seem difficult to achieve for a bubble deeply embedded in a dusty impact plume. Perhaps the radiation-condensation instability is restricted to the exposed edges of the plume, or occurs at times when the plume is more transparent---either early on when the vapor is hot and largely uncondensed, or late, when the plume is thinned out by expansion.
\item The mass of the bubble featured in Fig.~\ref{fig:the_one}, if compactified to solid density, would produce a planetesimal 9 m in radius. This is orders of magnitude too small compared to the sizes of ordinary (H, L, LL) and enstatite (EH) chondrite parent asteroids, estimated to be on the order of 100 km (e.g.~\citealt{miyamoto_etal_1982, edwards_blackburn_2020, trieloff_etal_2022}). The parent bodies of these classes of chondrite need to be this big to reproduce the most strongly thermally metamorphosed meteorites (types 6--7), from heat generated by $^{26}$Al at depth. Thus unless cavitation operated on scales vastly larger than we have imagined (perhaps possible with nebular shocks?), it cannot be the sole assembly mechanism for these chondrite classes; the small bodies created by cavitation would have to accrete en masse to form larger asteroids by another means---e.g.~pairwise accretion, or self-gravity, or a nebular gas-assisted mechanism such as the streaming instability (e.g.~\citealt{li_youdin_2021}). Note that this problem may not afflict CB/CH chondrites, and other types of carbonaceous chondrite, insofar as these classes are not thermally metamorphosed and therefore do not impose constraints on parent body size. Perhaps CB/CH chondrite parents are systematically much smaller than ordinary and enstatite chondrite parents.
\end{enumerate}

In the context of an impact plume, more multi-scale simulations like that of  \citet{stewart_lpsc6,stewart_lpsc7}, but including radiation (and dispensing with H$_2$, or including it with more realistic velocities), will be helpful. An alternative, semi-analytic approach would be to couple the equations governing the background plume (\citealt{choksi_etal_2021}) with our bubble evolution equations. 
In addition to pursuing radiative-hydrodynamic simulations and more formal stability analyses, we might also draw inspiration from fragmentation processes in other kinds of phase-changing, exploding or expanding media: supernovae; disk or stellar winds; and the expanding universe.

\vspace{0.2in}
\noindent 
We thank Sarah Stewart for a seminar that started this line of thinking, and Rixin Li and Siyi Xu for inspiring discussions. Erik Asphaug, Rick Binzel, Phil Carter, Jeffrey Fung, Emmanuel Jacquet, Anders Johansen, Myriam Telus, Andrew Youdin, and Shangjia Zhang provided extensive and thoughtful feedback on a draft manuscript that led to substantive changes. This work was supported by Berkeley's Esper Larsen, Jr. fund, and a Simons Investigator grant.

\software{numpy \citep{numpy_cite},
          scipy \citep{scipy_cite},
          matplotlib \citep{hunter_etal_2007}
          }






\bibliographystyle{aasjournal}
\bibliography{rci}



\end{document}